\documentclass[manuscript]{aastex}

\slugcomment{Submitted to Astrophysical Journal.}

\shorttitle{X-ray Line Feature at 3.5 keV}
\shortauthors{Phillips et al.}

\begin{document}


\title{THE X-RAY LINE FEATURE AT 3.5 keV IN GALAXY CLUSTER SPECTRA}


\author{K. J. H. Phillips\altaffilmark{1}, B. Sylwester\altaffilmark{2}, J. Sylwester\altaffilmark{2}}
\affil{$^1$ Earth Sciences Department, Natural History Museum, London SW7 5BD, UK}
\email{kennethjhphillips@yahoo.com}
\affil{$^2$ Space Research Center, Polish Academy of Sciences, Kopernika 11, 51-622 Wroc{\l}aw, Poland}
\email{bs@cbk.pan.wroc.pl,js@cbk.pan.wroc.pl}



\begin{abstract}
Recent work by Bulbul et al. and Boyarsky et al. has suggested that a line feature at $\sim 3.5$~keV in the X-ray spectra of galaxy clusters and individual galaxies seen with {\em XMM-Newton} is due to the decay of sterile neutrinos, a dark matter candidate. This identification has been criticized by Jeltema \& Profumo on the grounds that model spectra suggest that atomic transitions in helium-like potassium (K~{\sc xviii}) and chlorine (Cl~{\sc xvi}) are more likely to be the emitters. Here it is pointed out that the K~{\sc xviii} lines have been observed in numerous solar flare spectra at high spectral resolution with the RESIK crystal spectrometer and also appear in {\em Chandra} HETG spectra of the coronally active star $\sigma$~Gem. In addition, the solar flare spectra at least indicate a mean coronal potassium abundance which is a factor of between 9 and 11 higher than the solar photospheric abundance. This fact, together with the low statistical quality of the {\em XMM-Newton} spectra, completely accounts for the $\sim 3.5$~keV feature and there is therefore no need to invoke a sterile neutrino interpretation of the observed line feature at $\sim 3.5$~keV.
\end{abstract}


\keywords{galaxies: clusters: general --- galaxies: individual (M31) --- Sun: abundances --- Sun: X-rays, gamma rays --- X-rays: galaxies: clusters}



\section{Introduction}

In recent analyses of X-ray spectra of galaxy clusters, the Milky Way center, and individual galaxies including M31, \cite{bul14a} and \cite{boy14a} have claimed that an excess of emission over the range 3.47--3.51 keV (3.53--3.57~\AA) is a signature of the decay of sterile neutrinos (a dark matter candidate). The possibility that this line emission is due to helium-like potassium (\ion{K}{18}) was investigated using the AtomDB (APEC) atomic code but was rejected on the grounds that the observed line feature is about a factor 20 more intense than the potassium lines assuming the solar photospheric K abundance. This work has been criticized by \cite{jel15} who find that the line emission, which is significant at the 3 sigma level (though this is also disputed) is due to a combination of \ion{K}{18} and \ion{Cl}{17} line emission. A series of unpublished papers since then has continued the argument for and against the sterile neutrino identification \citep{bul14b,jel14,boy14b}, all based on interpretation of model spectra from the AtomDB code.

Unfortunately, these authors have neglected the fact that this region has been seen with high spectral resolution in numerous solar flare spectra with temperatures up to 22~MK (1.9~keV) using the RESIK crystal spectrometer \citep{jsyl05,jsyl10a} and in the spectra of a binary star system with an active corona ($\sigma$~Gem) using the {\em Chandra} HETG spectrometer \citep{hue13a,hue13b}. The RESIK observations have a spectral resolution of 11~eV (11.5~m\AA), and the {\em Chandra} HETG spectra a spectral resolution of 12~eV (12~m\AA). These are considerably better than the $\sim 100$~eV resolution of the MOS and PN CCD detectors on {\em XMM-Newton}. Consequently, the individual members of the \ion{K}{18} line complex ($w$, $x + y$, $z$ lines: notation of \cite{gab72}) at 3.473, 3.493, and 3.511~keV (photon energies based on wavelengths from \cite{kel87}) are well resolved in the solar flare spectra. Moreover, analysis of the well calibrated RESIK instrument allows absolute abundance determinations, and from 2795 individual spectra recorded during 20 flares between 2002 and 2003, \cite{jsyl10a} determined the mean K abundance (on a logarithmic scale with $A({\rm H}) = 12$) to be $A({\rm K}) = 5.86 \pm 0.23$ (s.d.), a factor of nearly 7 times the solar photospheric abundance of \cite{asp09} ($A({\rm K}) = 5.03 \pm 0.09$) or a factor of nearly 6 higher than that of \cite{caf11} (($A({\rm K}) = 5.11 \pm 0.09$). Such deviations of solar coronal abundances from photospheric abundances are well known and have been found to depend on the first ionization potential of an element (see Section 3).

In this paper, we extend the work of \cite{jsyl10a} by the addition to RESIK data-sets of several thousand spectra, with the consequence that our previous K abundance estimate for all RESIK flares is revised upwards. We then examine whether the spectral signature recorded by \cite{bul14a} and \cite{boy14b} are compatible with the \ion{K}{18} line emission based on these solar flare observations, and whether there is a need for invoking sterile neutrino decay.

\section{Observations}

\begin{figure*}
\epsscale{.80}
\plotone{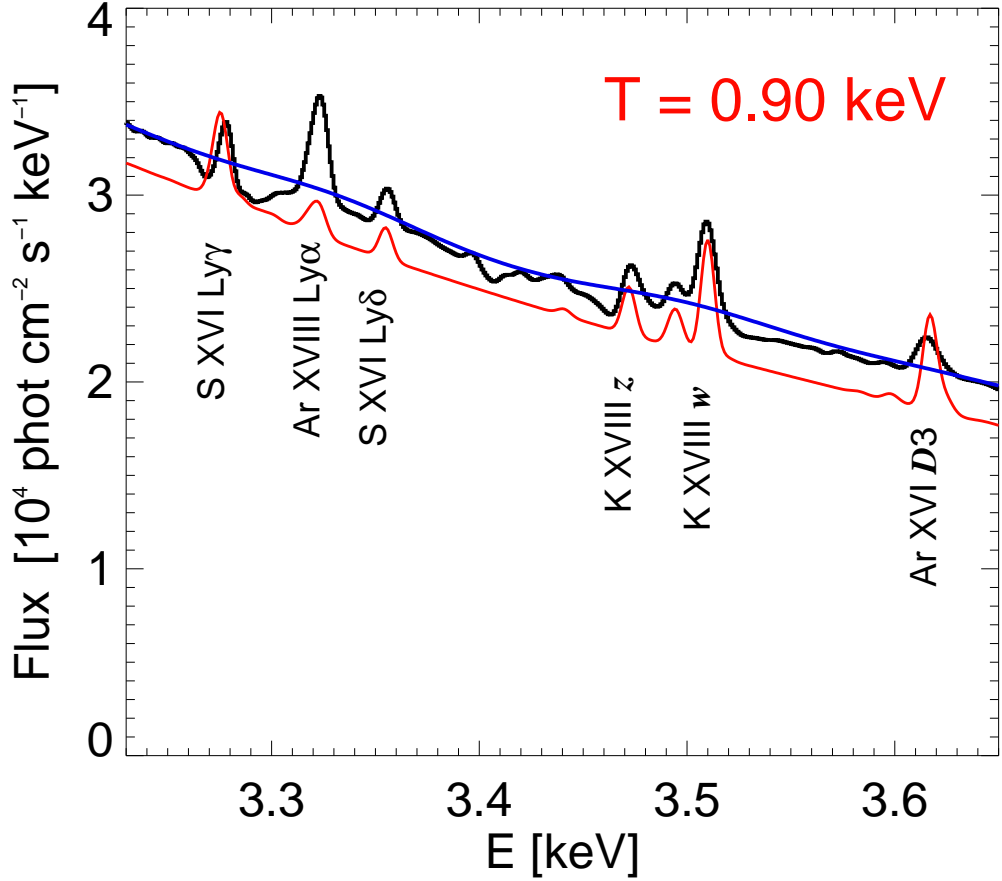}
\caption{Thick black curve: summed RESIK spectrum for the entire data-set (9295 spectra taken during 101 solar flares) in the channel~1 energy range (3.23--3.65~keV). The average temperature for this spectrum, estimated by the closest fit of a {\sc chianti} spectrum to the observed, is 0.90~keV (10.5~MK), somewhat higher than the averaged value of $T_{\rm GOES}$ (0.82~keV). Principal line features are identified by ion labels; more detailed information of lines occurring in this range are given in Table~\ref{line_ids}. Blue curve: RESIK spectrum convolved with a Gaussian profile with FWHM of 100~eV to match the resolution of the {\em XMM-Newton} detectors. Red curve: Theoretical spectrum from {\sc chianti} with temperature equal to 0.90~keV. Abundances from the {\sc chianti} ``coronal$\_$ext'' \citep{fel92b} abundance set are assumed.  \label{total_RESIK_sp}}
\end{figure*}

The RESIK instrument \citep{jsyl05} was a high-resolution crystal spectrometer on the {\em CORONAS-F} spacecraft, and was operational over the period 2002--2003 when solar activity was at a high level. The full spectral range of the instrument for an on-axis source, covered by four channels, was 2.05--3.69~keV (wavelength range 3.36--6.05~\AA). The spectral range was slightly extended for off-axis sources. Channel~1 (spectral range 3.26--3.69~keV) includes the He-like K (\ion{K}{18}) lines mentioned earlier as well H-like S (\ion{S}{16}) and Ar (\ion{Ar}{18}) Lyman lines and a dielectronic satellite line feature at 3.62~keV due to \ion{Ar}{16} made up of several lines but with one major contributor. Our earlier analysis \citep{jsyl10a} was based on 2795 spectra recorded in 20 flares. For this work, many additional flare spectra were added to a scientifically analyzable level (so-called Level~2: see http://www.cbk.pan.wroc.pl/experiments/resik/RESIK\_Level2/index.html), giving a total of 9295 spectra occurring during 101 flares. Figure~\ref{total_RESIK_sp} (black line) shows the summed spectrum of the entire data-set, so includes spectra over all development stages of the flares analyzed. The principal lines occurring in this range are identified by the emitting ions in this figure. Over-plotted in this figure (red lines) are synthetic spectra from the {\sc chianti} database and spectral code \citep{der97,lan12}. The observed spectra were convolved with a Gaussian profile having FWHM equal to 100~eV to match the energy resolution of the MOS and PN detectors on {\em XMM-Newton}; these are shown as the blue lines.

More details of the spectral lines of interest are given in Table~\ref{line_ids}. This table gives all significant lines occurring in the energy range of RESIK channel~1 as well as neighboring regions, including RESIK channels 2 to 4, for comparison with the lines listed by \cite{bul14a}. A few errors in the \cite{bul14a} list are corrected and photon energies (from the wavelengths of \cite{kel87}) and transitions added with estimates of temperature of maximum contribution functions (emissivities). Note that there are four \ion{K}{18} lines ($w$, $x$, $y$, and $z$) and not just two ($w$ and $z$) as stated by \cite{bul14a} and \cite{jel15}. Figure~\ref{total_RESIK_sp} shows that at the 100-eV resolution, the only recognizable line features above the continuum level are one due to the \ion{K}{18} lines and a second due to the combined \ion{S}{16} Ly$\gamma$ and \ion{Ar}{18} Ly$\alpha$ lines. The prominent \ion{Ar}{17} $w3$ line at 3.685~keV is on the high-energy edge of RESIK channel~1 and was not generally visible, although for flares with high spatial offset from the Sun's center this line sometimes fell within the observed channel~1 limits.

\begin{figure*}
\epsscale{.80}
\plotone{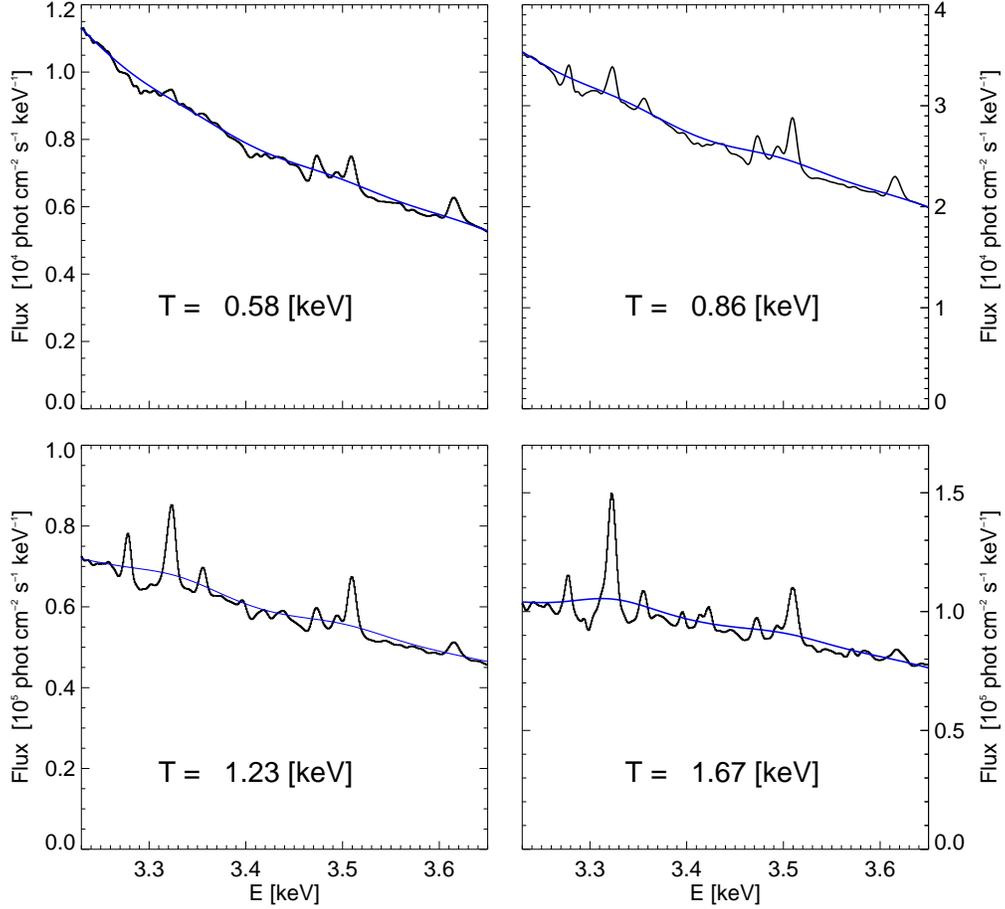}
\caption{Four solar flare spectra from RESIK channel 1 plotted against photon energy (keV) and summed over intervals of $T_{\rm GOES}$ (in keV); the means of the ranges are indicated in the legend (ranges are 0.34--0.69~keV; 0.69--1.03~keV; 1.12--1.46~keV; and 1.55--1.90~keV). Black lines: RESIK spectra; blue lines: RESIK spectra convolved with a gaussian having a FWHM of 100~eV, similar to the detectors on {\em XMM-Newton}. \label{RESIK_sp}}
\end{figure*}

Figure~\ref{RESIK_sp} shows RESIK channel~1 spectra extracted from the new data set and summed over narrow regions of temperature $T_{\rm GOES}$, taken to be that estimated from the flux ratio of the two X-ray channels of the {\em Geostationary Operational Environmental Satellites} ({\em GOES}\/). Our previous work has indicated that $T_{\rm GOES}$ is an accurate representation of temperature for the \ion{K}{18} lines as well as \ion{Ar}{17} lines seen in channel~2 \citep{jsyl10b}. The emission measures EM$_{\rm GOES}$ (emission measure = $N_e^2 V$ where $N_e = $ electron density and $V =$ emitting volume) for each RESIK spectrum were estimated from the ratio of the observed spectral flux to the theoretical spectral flux from the {\sc chianti} code at the temperature $T_{\rm GOES}$. The vertical scale in Figure~\ref{RESIK_sp} is flux (photon cm$^{-2}$ s$^{-1}$ keV$^{-1}$) normalized to a volume emission measure  of $10^{48}$~cm$^{-3}$.  Through a careful adjustment of pulse-height analyzers during the {\em CORONAS-F\/} mission, crystal fluorescence background emission was practically eliminated for RESIK channels~1 and 2, so the background levels indicated for each spectrum of Figure~\ref{RESIK_sp} are solar continuum. With the exception of the \ion{Ar}{16} satellite feature at 3.62~keV, the lines are progressively stronger with increasing $T_{\rm GOES}$ and the continuum slope is progressively flatter with increasing $T_{\rm GOES}$.

The spectral resolution of RESIK channel~1 is such that the \ion{K}{18} $w$, $x+y$, and $z$ lines are apparent as three separate line features. We cannot state with certainty whether RESIK sees the \ion{Cl}{17} Ly$\beta$ line (3.507~keV), mentioned as possibly present in {\em XMM-Newton} spectra of galaxy clusters and individual galaxies \cite{jel15}. The \ion{Cl}{17} Ly$\alpha$ line, which should be more intense, occurs at the low-energy edge of RESIK channel~2 and is blended with a dielectronic satellite of \ion{S}{14}; the blended feature has been observed in the spectra of some flares with high spatial offsets but the line flux is uncertain because of its vicinity to the end of the crystal range. However, the \ion{Cl}{16} $1s^2 - 1s2l$ ($l=s$, $p$) lines at $\sim 2.79$~keV have been recorded in channel~3 during many flares and from which a solar flare chlorine abundance \citep{bsyl11} has been estimated ($A({\rm Cl}) = 5.75 \pm 0.26$), compared with $A({\rm Cl}) = 5.5$ from an early analysis of HCl sunspot spectra by \cite{hal72} (see remarks on the latter determination by \cite{asp09}).  Other lines are expected in the region of the \ion{K}{18} lines, in particular \ion{K}{17} dielectronic satellites, the fluxes of which were calculated by \cite{jsyl10a}. The most significant include satellites $j$ and $k$ (3.474, 3.477~keV respectively), which are blended with \ion{K}{18} line $z$, and $d13$ and $d15$ (3.503~keV) which are blended with \ion{K}{18} line $w$. At temperatures $< 1$~keV, the fluxes of $j$ and $k$ are calculated to be comparable to \ion{K}{18} line $z$ but decreasing for higher temperatures (the satellite-to-line $w$ ratio has an approximately $T^{-1}$ dependence).

The greatly extended RESIK channel~1 data set has allowed us to refine the determination of the solar flare K abundance adopting the same method used in our previous analyses. For each spectrum, continuum fluxes in units of photon cm$^{-2}$ s$^{-1}$ bin$^{-1}$ on either side of the \ion{K}{18} line group were measured and an average value calculated; in this spectral region, a RESIK energy bin is approximately 1.54~eV (1.56~m\AA). The energy intervals containing only continuum were 3.462--3.465~keV and 3.529--3.543~keV. The total flux $F_{\rm K XVIII}$ in the \ion{K}{18} $w$, $x+y$, and $z$ lines together with continuum was then determined over the energy interval 3.465--3.520~keV. With $N_1$, $N_2$, and $N_3$ the number of RESIK energy bins in the 3.462--3.465~keV, 3.529--3.543~keV, and 3.465--3.520~keV, the total flux in the \ion{K}{18} $w$, $x+y$, and $z$ lines is therefore
\begin{equation}
F_{\rm K\, XVIII} = F(3.465 - 3.520) - N_3 \times C_{\rm av}
\end{equation}
\noindent where $C_{\rm av}$ is
\begin{equation}
C_{\rm av} = \frac{1}{2} \times [ F(3.462-3.465) / N_1 + F(3.529-3.543) / N_2 ].
\end{equation}

\begin{figure*}
\epsscale{.80}
\plotone{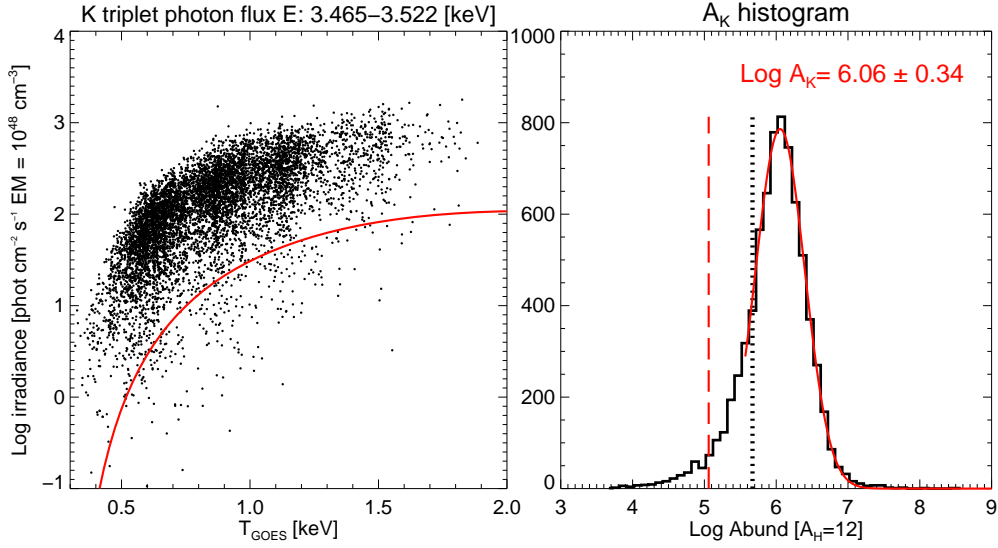}
\caption{Left panel: Points show the flux in the energy interval 3.465--3.520~keV with averaged flux in two narrow neighboring line-free spectral ranges subtracted. Solid red curve: $G(T)$ function for \ion{K}{18} lines $w$, $x$, $y$, and $z$ lines at 3.465--3.520~keV from {\sc chianti} in photons cm$^{-2}$ s$^{-1}$ for a volume emission measure of $10^{48}$ cm$^{-3}$ and solar photospheric abundance \citep{asp09}. The points represent estimated \ion{K}{18} line fluxes (photon cm$^{-2}$ s$^{-1}$) divided by {\em GOES} emission measure EM$_{\rm GOES}$ in cm$^{-3}$ for 7439 RESIK spectra (the remaining 1856 spectra in the complete sample have \ion{K}{18} line fluxes too weak to analyze for this purpose) and multiplied by $10^{48}$. Right panel: Estimates of the solar flare K abundance plotted as a frequency histogram. The best-fit Gaussian to the range indicated by the red curve gives $A({\rm K}) = 6.06 \pm 0.34$ (s.d.). The vertical lines are (dashed red) photospheric K abundance \citep{asp09} and (dotted black) from the {\sc chianti} ``coronal$\_$ext'' \citep{fel92b} abundance set. \label{goft}}
\end{figure*}

\noindent Note that the very slight contributions made by the \ion{S}{16} high-$n$ lines and \ion{Cl}{17} line in the vicinity of the \ion{K}{18} lines were neglected. Using the temperature and emission measure obtained from {\em GOES} channel ratios as earlier described, the irradiances of the \ion{K}{18} lines were found from
\begin{equation}
I_{\rm K\,XVIII} = \frac{F_{\rm K\, XVIII}}{{\rm EM}_{\rm GOES}} \times 10^{48}.
\end{equation}
\noindent These irradiances are plotted for each spectrum analyzed against $T_{GOES}$ in Figure~\ref{goft} (left panel). As expected, the distribution of points has the same temperature dependence as the theoretical contribution function, calculated from {\sc chianti}, for the \ion{K}{18} lines using the photospheric K abundance of $A({\rm K}) = 5.03$ \citep{asp09} shown as the red curve. The scatter of points reflects the weakness of the \ion{K}{18} lines in RESIK spectra, but it is clear from the displacement of the points above the red curve that the flare abundance is considerably larger than the photospheric. An estimate of the abundance RESIK spectra is possible through the deviation of each plotted point in Figure~\ref{goft} (left panel) from the theoretical contribution function; the distribution of these estimates against the estimated value of $A({\rm K})$ is shown as a histogram plot in Figure~\ref{goft} (right panel). The best-fit Gaussian to the distribution is also shown, the peak and width of which give our estimate of the solar flare K abundance, $A({\rm K}) = 6.06 \pm 0.34$ (s.d.). This should be compared with the result from our previous smaller sample of solar flares (Figure 3 of \cite{jsyl10a}), in which we derived $A({\rm K}) = 5.86 \pm 0.23$. The present abundance estimate is a factor of approximately 11 higher than the photospheric abundance of \cite{asp09}.

Our estimate of the solar flare abundance relies heavily on the atomic data for helium-like potassium in the {\sc chianti} database, in particular the values of the effective collision strengths which, in the absence of specific calculations available for the release of {\sc chianti} v.~7, are based on interpolations of helium-like ions of other elements. Since errors in these values folds directly into the abundance determinations, it is advisable to check the accuracy of the atomic data wherever possible. Since our earlier work \citep{jsyl10a}, new calculations have appeared \citep{agg12} based on the {\sc darc} (Dirac Atomic R-matrix Code) for \ion{K}{18} and other He-like ions. We compared the effective collision strengths for the principal atomic transitions involved in the \ion{K}{18} $w$, $x$, $y$, and $z$ lines finding, for the temperature range of interest ($T_{\rm GOES} > 0.4$~keV or $>4.6$~MK), only a few percent differences. This vindication of the atomic data used in our analysis based on {\sc chianti} thus adds confidence to our abundance estimates.

\section{Discussion and Conclusions}

Using a much extended data-base of solar flare spectra (spectral resolution 11~eV) from the RESIK instrument on {\em CORONAS-F} spacecraft over our earlier analysis \citep{jsyl10a}, our estimates of the \ion{K}{18} line irradiance lead to a solar flare abundance of K of $A({\rm K})= 6.06 \pm 0.34$ or a factor of 11 more than the photospheric abundance of \cite{asp09} ($A({\rm K}) = 5.03 \pm 0.09$) or a factor 9 more than that of \cite{caf11} ($A({\rm K}) = 5.11 \pm 0.09$). Our present abundance estimate is slightly more than was found in an earlier, less complete analysis of RESIK spectra ($A({\rm K}) = 5.86 \pm 0.23$: \cite{jsyl10a}). It may also be compared with the value $A({\rm K})= 5.7$ adopted from analysis of a \ion{K}{9} far-ultraviolet line observed with the Coronal Diagnostic Spectrometer on {\em Solar and Heliospheric Observatory} \citep{lan02}. Although the \ion{K}{18} lines are not very strong in RESIK spectra, the plot in Figure~\ref{goft} (left panel) clearly shows the enhanced abundance of K. In addition, for the RS CVn binary star system $\sigma$ Gem observed by {\em Chandra}, the K abundance was found to be a factor 4 more than solar photospheric \citep{hue13a}. Such abundance anomalies are well documented for the solar corona, particularly solar flare plasmas, and are commonly attributed to the ``FIP effect'' because of an apparent dependence on the value of the first ionization potential (FIP) of element abundances in the solar corona and solar wind particles \citep{mey85,fel92b,fel00}: low-FIP elements (i.e., FIP $\lesssim 10$~eV) are generally observed to have coronal abundances enhanced with respect to photospheric but high-FIP elements are not. The exact nature of the FIP effect in flare plasmas is still being actively discussed by several authors. For Si (FIP = 8.2~eV), a recent analysis of RESIK observations \citep{bsyl15} indicate no enhancement over photospheric, while for Fe (FIP = 7.9~eV), \cite{den15} from X-ray spectra found only a modest enhancement over photospheric and \cite{war14} using extreme ultraviolet spectra found none at all. This possibly indicates that the boundary between high-FIP and low-FIP elements is less than 10~eV. Comparison with element abundances from solar energetic particle observations (e.g. \cite{rea14,zur02}) is more problematic. First, impulsive particle events show large time variations and most analyses eliminate such events from averaged abundance values. Also, recovery of abundances at the coronal sources of particle events must take account of differential transport effects which are considerable. With these and other provisos in mind, \cite{rea14} finds that averaged abundance estimates for 54 gradual solar particle events show a K abundance enhancement over the \cite{caf11} photospheric abundance of only 2 (range 1.7--3). A difference in the nature of the FIP effect for solar energetic particles in interplanetary space and those from X-ray and ultraviolet observations, which refer to closed magnetic structures in the solar corona, is thus indicated. All these findings are relevant to theoretical mechanisms for the FIP effect which have been discussed by \cite{hen98} and \cite{lam12}.

The $\sim 3.5$~keV spectral region has been observed with {\em XMM-Newton} by \cite{bul14a} in 73 galaxy clusters and by \cite{boy14a} in the Perseus galaxy cluster and the Andromeda galaxy M31. The line feature noted by \cite{bul14a} at 3.47--3.51~keV has $>3 \sigma$ significance in stacked {\em XMM-Newton} spectra, though this is disputed by \cite{jel15}. The interval over which the emission is enhanced, 3.55--3.57~keV, is very close to that of the energy range of the \ion{K}{18} $w-z$ line emission. \cite{bul14a} analyze the {\em XMM-Newton} data using the AtomDB atomic physics package with solar photospheric abundances from \cite{and89}. More recent and comprehensive lists of photospheric abundances are given by \cite{asp09} and \cite{caf11}, with minor revisions to the abundances of elements relevant to the work of \cite{bul14a} and \cite{jel15}. Using $A({\rm K})= 5.13$ (given as $5.03 \pm 0.09$ by \cite{asp09} and $5.11 \pm 0.09$ by \cite{caf11}), \cite{bul14a} dismiss the identification of the 3.47--3.51~keV emission feature as \ion{K}{18} lines on the grounds that fits with the AtomDB package assuming a collisional ionization and excitation model require a K abundance of a factor of $\sim 20$ over the solar photospheric value (although this depends on the temperature model that they use), and instead attribute the feature to the decay of sterile neutrinos, a dark matter candidate, with mass 7.1~keV. The factor $\sim 20$ that \cite{bul14a} state the photospheric abundance of K needs to be enhanced is derived from a temperature model using a fit to the continuum from APEC in the XSPEC spectroscopy modeling procedure, with temperatures between 6~keV and 10~keV. However, these high temperatures disagree with the observed ratio of \ion{Ca}{20} to \ion{Ca}{19} lines \citep{jel14}.

Clearly, the case for identifying the 3.55--3.57~keV feature to sterile neutrino decay is much weakened if the X-ray emission sources observed by {\em XMM-Newton} have an origin like solar flare plasmas for which K is enhanced by some mechanism. Much of the emission from galaxy clusters arises from intra-cluster thermal plasmas in collisional ionization and excitation equilibrium with temperatures in the range 1--10~eV \citep{sar03}. If a magnetic mechanism like a ponderomotive force as proposed by \cite{lam12} for solar coronal plasmas is involved in the origin of this hot gas, then enhancements of K could result. For the case of individual galaxies like M31, the X-ray emission above 2~keV, with luminosities of $10^{39}$--$10^{40}$~erg~s$^{-1}$, is mostly from spatially unresolved high-luminosity low-mass X-ray binary systems \citep{tak04} with combined estimated X-ray luminosity of up to $\sim 5 \times 10^{39}$ erg~s$^{-1}$; no single black hole or similar accretion source is dominant or even significant. At softer energies, there is a diffuse emission attributed to thermal plasma with temperatures $<1$~keV. If the origin of the X-ray emission from the X-ray binaries is a ponderomotive force associated with Alfv\'{e}n waves, then abundance enhancements in K like those in solar flare plasmas could arise if the gas being accreted by the evolved component of the binary system has its origin in the outer atmosphere of the donor (unevolved) star. The K abundance enhancement estimated here from RESIK solar flare spectra only needs a factor  of $\sim 1.8$ to explain the factor 20 mentioned by \cite{bul14a}, and this could easily be explained by the 100-eV spectral resolution and the very low statistical significance of their signal, even in the summed spectra of the galaxy clusters that they observed with {\em XMM-Newton}. This could also explain the slight difference in the energy of the observed feature (3.55--3.57~keV) and the range of the \ion{K}{18} lines (3.47--3.51~keV).

In addition to the \ion{K}{18} lines, small contributions may be made by the \ion{Cl}{17} Ly$\beta$ line at 3.507~keV, very near the \ion{K}{18} $w$ line, as noted by \cite{jel15}. To estimate this contribution, we use RESIK observations of the \ion{Cl}{16} $w$, $x+y$, and $z$ lines (2.76--2.79~keV) from which a chlorine abundance $A({\rm Cl}) = 5.75 \pm 0.26$ was derived \citep{bsyl11}. This is consistent with the estimate of \cite{hal72} from a sunspot HCl spectrum, $A({\rm Cl}) = 5.5 \pm 0.3$, as is expected from the solar coronal FIP effect with the high value of chlorine's FIP (12.97~eV). Using the RESIK solar flare abundance estimates of Cl and K (from this work), we estimate that only about 2\% of the 3.55--3.57~keV emission is due to Cl at a temperature of 2~keV, but increasing to 13\% for 5~keV (one of the temperatures given by \cite{bul14a}). Thus, we confirm the calculation made by \cite{jel15} that Cl does not seem to be a dominant contributor to the \ion{K}{18} line emission.

High-$n$ hydrogen-like S (\ion{S}{16}) lines occur in the 3.55--3.57~keV region, as indicated by Table~\ref{line_ids}, in particular the series limit of these lines occurs between the \ion{K}{18} $w$ and $x+y$ lines in RESIK solar flare spectra. Figure~\ref{total_RESIK_sp} and the highest-temperature panel of Figure~\ref{RESIK_sp} certainly show the \ion{S}{16} Ly$\delta$ and Ly$\epsilon$ lines (3.355~keV and 3.398~keV) with possible evidence of higher-$n$ members, but it is evident that their contribution to the \ion{K}{18} lines is insignificant.

In summary, solar flare spectra from the RESIK instrument on {\em CORONAS-F} indicate a mean coronal potassium abundance which is a factor of between 9 and 11 higher than the solar photospheric abundance. For solar coronal plasmas, some mechanism operates to enhance the coronal abundances of elements with low first ionization potential, possible ones being discussed by \cite{hen98} and \cite{lam12}. However, the boundary between high-FIP and low-FIP elements, previously cited to be 10~eV, is not completely established, recent X-ray and extreme ultraviolet spectroscopic flare observations indicating that Si and Fe (FIP = 8.2~eV, 7.9~eV respectively) are either not enhanced at all or are only slightly enhanced over photospheric. If this or some equivalent mechanism applies to the stellar component of X-ray emission in individual galaxies or the high-temperature gas in galaxy clusters, the observed line feature at 3.55--3.57~keV is readily explained, particularly in view of the low statistical quality of the X-ray spectra from {\em XMM-Newton} obtained by \cite{bul14a}, and thus a sterile neutrino interpretation of the observed line feature at $\sim 3.5$~keV is unnecessary.

\acknowledgments

We acknowledge financial support from the Polish National Science Centre grant number UMO-2013-11/B/ST9/00234. BS and JS are grateful for the support from the International Space Science Institute in Bern, Team No. 276 (``Non-­-Equilibrium Processes in the Solar Corona and their Connection to the  Solar Wind"). The {\sc chianti} atomic database and code is a collaborative project involving George Mason University, University of Michigan (USA), and University of Cambridge (UK). We are grateful to the {\sc chianti} team for supplying us with \ion{S}{16} high-$n$ line data.

{\em Facilities:} \facility{GOES}, \facility{CORONAS/RESIK}

\bibliographystyle{apj}

\bibliography{RESIK}

\newpage
\begin{deluxetable}{lllccc}
\tabletypesize{\tiny}
\tablecaption{Lines contributing to the 2.0--4.1~keV (3.0--6.2~\AA) solar X-ray spectra \label{line_ids}}
\tablewidth{0pt}
\tablehead{\colhead{Ion}&\colhead{Line notation} & \colhead{Transition} & \colhead{Energy (keV)$^a$} & \colhead{$T_{\rm max}$ (keV)$^b$}  & Seen by RESIK (Y/N)\\}
\startdata
Si XIV & Ly$\alpha$ & $1s\,^2S_{1/2} - 2p\,^2P_{1/2,3/2}$ & 2.004, 2.006 & 1.4  \\
Al XIII & Ly$\beta$ & $1s\,^2S_{1/2} - 3p\,^2P_{1/2,3/2}$ & 2.048 & 1.1 \\
\\
RESIK Channel 4 & (2.05--2.48 keV) \\
Si XII & $D4$ & $1s^2\,2p\,^2P_{3/2} - 1s2p(^2P)\,4p\,^2D_{5/2}$ & 2.228 & 0.8 & Y$^c$ \\
Si XIII & $w3$ & $1s^2\,^1S_0 - 1s3p\,^{1}P_1$ & 2.183 & 1.0 & Y \\
Si XIII & $w4$ & $1s^2\,^1S_0 - 1s4p\,^{1}P_1$ & 2.294 & 1.0 & Y \\
Si XIII & $w5$ & $1s^2\,^1S_0 - 1s5p\,^{1}P_1$ & 2.346 & 1.0 & Y \\
Si XIV & Ly$\beta$ & $1s\,^2S_{1/2} - 3p\,^2P_{1/2,3/2}$ & 2.376, 2.377 & 1.4 & Y \\
S XV & $z$ & $1s^2\,^1S_0 - 1s2s\,^3S_1$ & 2.431 & 1.4 & Y \\
S XV & $x+y$ & $1s^2\,^1S_0 - 1s2p\,^3P_{1,2}$ & 2.447 & 1.4 & Y \\
S XV & $w$ & $1s^2\,^1S_0 - 1s2p\,^1P_1$ & 2.461 & 1.4 & Y \\
Si XIV & Ly$\gamma$ & $1s\,^2S_{1/2} - 4p\,^2P_{1/2,3/2}$ & 2.506 & 1.4 & Y \\
\\
RESIK Channel 3 & (2.55--2.85 keV) \\
Si XIV & Ly$\delta$ & $1s\,^2S_{1/2} - 5p\,^2P_{1/2,3/2}$ & 2.566 & 1.4 & Y \\
S XVI & Ly$\alpha$ & $1s\,^2S_{1/2} - 2p\,^2P_{1/2,3/2}$ & 2.620, 2.623 & 2.2 & Y \\
Cl XVI & $z$ & $1s^2\,^1S_0 - 1s2s\,^3S_1$ & 2.757 & 1.5 & Y$^d$ \\
Cl XVI & $x+y$ & $1s^2\,^1S_0 - 1s2p\,^3P_{1,2}$ & 2.775 & 1.5 & Y$^d$ \\
Cl XVI & $w$ & $1s^2\,^1S_0 - 1s2p\,^1P_1$ & 2.790 & 1.5 & Y$^d$ \\
S XIV & $D3$ & $1s^2\,2p\,^2P_{3/2} - 1s2p(^2P)\,3p\,^2D_{5/2}$ & 2.817 & 1.1 & Y \\
S XV & $w3$ & $1s^2\,^1S_0 - 1s3p\,^{1}P_1$ & 2.884 & 1.4 & Y \\
\\
RESIK Channel 2 & (2.90--3.24 keV)\\
S XIV & $D4$ & $1s^2\,2p\,^2P_{3/2} - 1s2p(^2P)\,4p\,^2D_{5/2}$ & 2.957 & 1.1 &  Y$^e$ \\
Cl XVII & Ly$\alpha$ & $1s\,^2S_{1/2} - 2p\,^2P_{1/2,3/2}$ & 2.958, 2.963 & 2.4 &  Y$^{e}$? \\
S XV & $w4$ & $1s^2\,^1S_0 - 1s2p\,^{1}P_1$ & 3.033 & 1.4 &  Y \\
S XV & $w5$ & $1s^2\,^1S_0 - 1s5p\,^{1}P_1$ & 3.101 & 1.4 &  Y \\
Ar XVII & $z$ & $1s^2\,^1S_0 - 1s2s\,^3S_1$ & 3.104 & 1.9 &  Y$^e$\\
S XVI & Ly$\beta$ & $1s\,^2S_{1/2} - 3p\,^2P_{1/2,3/2}$ & 3.107 & 2.2 &  N$^e$?\\
Ar XVII & $x+y$ & $1s^2\,^1S_0 - 1s2p\,^3P_{1,2}$ & 3.124 & 1.9 &  Y\\
Ar XVII & $w$ & $1s^2\,^1S_0 - 1s2p\,^1P_1$ & 3.140 & 1.9 &  Y \\
S XV & $w6$ & $1s^2\,^1S_0 - 1s5p\,^{1}P_1$ & 3.140 & 1.4 &  \\
\\
RESIK Channel 1 & (3.26--3.69 keV)\\
Cl XVI & $w3$ & $1s^2\,^1S_0 - 1s3p\,^{1}P_1$ & 3.272 & 1.5 &  N \\
S XVI & Ly$\gamma$ & $1s\,^2S_{1/2} - 4p\,^2P_{1/2,3/2}$ & 3.277 & 2.2 & Y\\
Ar XVIII & Ly$\alpha$ & $1s\,^2S_{1/2} - 2p\,^2P_{1/2,3/2}$ & 3.323 & 3.1 &  Y\\
S XVI & Ly$\delta$ & $1s\,^2S_{1/2} - 5p\,^2P_{1/2,3/2}$ & 3.355 & 2.2 &  Y\\
S XVI & Ly$\epsilon$ & $1s\,^2S_{1/2} - 6p\,^2P_{1/2,3/2}$ & 3.398 & 2.2 &  Y \\
S XVI & Ly$\zeta$--Ly$\chi$ & $1s-7p$ to $1s-10p$ & 3.423--3.459 & 2.2 &  Y? \\
K XVIII & $z$ & $1s^2\,^1S_0 - 1s2s\,^3S_1$ & 3.472 & 2.2 &  Y \\
K XVIII & $x+y$ & $1s^2\,^1S_0 - 1s2p\,^3P_{1,2}$ & 3.493 & 2.2 &  Y\\
S XVI & Series limit & $1s-\infty$ & 3.494 & -- &  N?\\
Cl XVII & Ly$\beta$ & $1s\,^2S_{1/2} - 3p\,^2P_{1/2,3/2}$ & 3.507 & 2.4 & N \\
K XVIII & $w$ & $1s^2\,^1S_0 - 1s2p\,^1P_1$ & 3.510 & 2.2 &  Y\\
Ar XVI  & $D3$ & $1s^2\,2p\,^2P_{3/2} - 1s2p(^2P)\,3p\,^2D_{5/2}$ & 3.615 & 1.5 & Y\\
Ar XVII & $w3$ & $1s^2\,^1S_0 - 1s3p\,^{1,3}P_1$ & 3.685 & 1.9  & Y\\
\\
K XIX & Ly$\alpha$ & $1s\,^2S_{1/2} - 2p\,^2P_{1/2,3/2}$ & 3.699, 3.704 & 3.8  \\
Ar XVI & $D4$ & $1s^2\,2p\,^2P_{3/2} - 1s2p(^2P)\,4p\,^2D_{5/2}$ & 3.794 & 1.5  \\
Ca XIX & $z$ & $1s^2\,^1S_0 - 1s2s\,^3S_1$ & 3.861 & 2.4  \\
Ar XVII & $w4$ & $1s^2\,^1S_0 - 1s4p\,^{1}P_1$ & 3.875 & 1.9  \\
Ca XIX & $y$ & $1s^2\,^1S_0 - 1s2p\,^3P_{1}$ & 3.884 & 2.4 \\
Ca XIX & $x$ & $1s^2\,^1S_0 - 1s2p\,^3P_{2}$ & 3.888 & 2.4 \\
Ca XIX & $w$ & $1s^2\,^1S_0 - 1s2p\,^1P_{1}$ & 3.903 & 2.4 \\
Ar XVIII & Ly$\beta$ & $1s\,^2S_{1/2} - 3p\,^2P_{1/2,3/2}$ & 3.936 & 3.1  \\
Ar XVII & $w5$ & $1s^2\,^1S_0 - 1s5p\,^{1}P_1$ & 3.964 & 1.9 \\
Ca XX & Ly$\alpha$ & $1s\,^2S_{1/2} - 2p\,^2P_{1/2,3/2}$ & 4.100, 4.108 & 4.3  \\
\enddata

\tablenotetext{a}{Except for Si XII and Cl XVI lines, line energies are from wavelengths given by \cite{kel87}.}
\tablenotetext{b}{Approximate values: maximum temperature not well defined as contribution function only slowly declines with $T$.}
\tablenotetext{c}{See \cite{phi06}; this line predominates but there are many other Si XII lines in this feature. This applies to other dielectronic lines (e.g. the Ar XVI $D3$ line at 3.62~keV) with similar transitions. }
\tablenotetext{d}{See \cite{bsyl11}.}
\tablenotetext{e}{.Lines blended in RESIK spectra.}

\end{deluxetable}

\end{document}